\documentclass[sigconf, nonacm]{acmart}
\usepackage{CJKutf8}
\AtBeginDocument{%
  \providecommand\BibTeX{{%
    \normalfont B\kern-0.5em{\scshape i\kern-0.25em b}\kern-0.8em\TeX}}}
\usepackage{multirow}

\usepackage{array}
\usepackage{booktabs}
\usepackage{setspace}
\begin{document}
\begin{CJK}{UTF8}{gbsn}

\title{A Concept Knowledge-Driven Keywords Retrieval Framework for Sponsored Search}

\author{Yijiang Lian}
\email{lianyijiang@baidu.com}
\affiliation{%
  \institution{Baidu}
}

\author{Yubo Liu}
\email{liuyubo01@baidu.com}
\affiliation{%
  \institution{Baidu}
}

\author{Zhicong Ye}
\email{yezhicong@baidu.com}
\affiliation{%
  \institution{Baidu}
}

\author{Liang Yuan}
\email{yuanliang@baidu.com}
\affiliation{%
  \institution{Baidu}
}

\author{Yanfeng Zhu}
\email{zhuyanfeng@baidu.com}
\affiliation{%
  \institution{Baidu}
}

\author{Min Zhao}
\email{zhaomin@baidu.com}
\affiliation{%
  \institution{Baidu}
}

\author{Jianyi Cheng}
\email{chengjianyi@baidu.com}
\affiliation{%
  \institution{Baidu}
}

\author{Xinwei Feng}
\email{fengxinwei@baidu.com}
\affiliation{%
  \institution{Baidu}
}

\renewcommand{\shortauthors}{}


\begin{abstract}
  In sponsored search, retrieving synonymous \emph{keywords} for \emph{exact match}
  type is important for accurately targeted advertising.
  Data-driven deep learning-based method has been proposed to tackle this problem. An apparent disadvantage of this method is its
  poor generalization performance on entity-level long-tail instances,
  ev- en though they might share similar concept-level patterns with frequent instances. With the help of a large knowledge base, we find that
  most commercial synonymous query-\emph{keyword} pairs can be abstracted into meaningful conceptual patterns through concept tagging.
  Based on this fact, we propose a novel knowledge-driven conceptual retrieval framework to
  mitigate this problem,
  which consists of three parts:
  data conceptualization, matching via conceptual patterns and concept-augmented discrimination.
  Both offline and online experiments show that our method is very effective.
  This framework  has been successfully
  applied to Baidu's sponsored search system, which yields a
  significant improvement in revenue.

\end{abstract}


\keywords{Sponsored Search; Keyword Matching; Keyword Retrieval;
Paraphrase Pattern; Knowledge Graph}

\maketitle
\section{Introduction}

Sponsored search advertising refers to the placement of ads on
search result pages above or next to organic search results.
As these sponsored ads directly target user's query intention,
they usually have a much higher conversion ratio.
In the past few years, search advertising has become one of
the most popular forms of digital advertising worldwide.

One of the most critical modules of the sponsored search system
is the \emph{keyword} matching module, which is to match users' queries to advertisers' bidding \emph{keywords}.
(\emph{Keyword} is utilized to represent queries purchased by the advertisers
in particular).
Mainstream search engine companies provide a structured bidding language, with which the advertisers
can specify how would their purchased keywords be matched to the online queries.
Generally, there are three \emph{match types}
provided \footnote{https://support.google.com/google-ads/answer/7478529?hl=en}:
\emph{exact, phrase, broad}.
Under the \emph{exact match type}, ads would be
eligible to appear when a user searches for the specific \emph{keyword} or its synonymous variants.
For \emph{phrase match type}, the matched queries should contain the \emph{keyword} or the synonymous variants of the \emph{keyword}.
\emph{Broad match type} further relaxes the matching restrictions to the semantic relevance level.
Due to the targeting of precise traffic, most customers prefer the \emph{exact match type}, which accounts for a large portion of the \emph{keyword} revenue in most search engine companies.
In this article, we pay more attention to the problem of synonymous matching under the \emph{exact match type}.



Since the synonymous query-\emph{keyword} relationship is quite scar- ce,
while the volumes of the industrial queries and \emph{keywords} are extremely huge,
the traditional boolean retrieval framework, which works well in \emph{broad match} scenarios,
is quite inefficient in this \emph{exact match} scenario \cite{lian2020para}.
Recently, data-driven deep learning methods have been applied in this scenario \cite{lian2020para}, where a translation model is trained on a high-quality paraphrasing dataset and is used to make a generalization to link more synonymous queries and keywords.

One problem of the data-driven deep learning-based approach is
its poor generalization performance on the long-tail instances \cite{hendrycks2020pretrained}.
Considering that nearly 70\% of the queries are entity related \cite{guo2009named},
we especially focus on those entity-level long-tail cases.
For example, if  \emph{double-fold eyelid operation} and \emph{Los Angeles} are commonly observed in the training data,
the translation model can generate lots of high quality paraphrases for queries
like \emph{the price of double-fold eyelid operation in Los Angeles}; however, it would fail on queries like
\emph{the price of liposuction in Denver} if  \emph{liposuction} and  \emph{Denver} are rare in the training data, even though
these two queries share a same abstract conceptual pattern \emph{the price of [aesthetic sur-}
\emph{gery] in [location]}.
And the same is true for the discriminant model.
With a limited number of synonymous training data in hand and a huge number of queries to be addressed,
long-tail cases should be regularly encountered for the industrial models.
How to make our retrieval system robust on these long-tail cases becomes an important and urgent problem to be addressed.

Conceptual abstraction and reasoning are common in the human learning process. Given a synonymous training instance
(\emph{how much does double-fold eyelid operation cost in Los Angeles
= the price of double-fold eyelid operation in Los Angeles}),
if we know that \emph{double-fold eyelid operation} belongs to the concept of aesthetic surgery,
and \emph{Los Angeles} belongs to the concept of \emph{[location]},
we can easily abstract the original paraphrase instance into a conceptual pattern form
(\emph{how much does [aesthetic surgery] cost in [location]
= the price of  [aesthetic surgery] in [location]}).
When a long-tail query like \emph{the price of liposuction in Denver} comes,
if we get the knowledge that \emph{liposuction} belongs to \emph{[aesthetic
surgery]} and it has an alias "lipo",
we can effortlessly inference out a new paraphrase instance:
(\emph{how much does lipo cost in Denver = the price of  liposuction in Denver}). In this way,
the paraphrasing capability has been transferred onto the long-tail instances.





Inspired by this idea, we propose a concept knowledge-driven \emph{keyword} retrieval framework,
which comprises three phases: data conceptualization,
matching via conceptual patterns, and concept-augmented discrimination.
In the first phase of data conceptualization, the original paraphrasing data
is transformed into the conceptual pattern form by concept tagging, where each token
in the sentence would be labeled with a concept label.
Core commercial concepts in the pattern are selected for further
generalization usage and others are ignored.
Secondly, a deep neural translation model is directly trained on this conceptualized data to capture the synonymous pattern expression's variations. This
model is used to link query patterns with \emph{keyword} patterns.
Finally, a concept-augmented discrimination model is utilized to filter out nonsynonymous cases.

Our offline experiments revealed that this conceptual retrieval framework can significantly
boost the retrieval performance for long-tail cases.
Besides, this framework has been successfully applied in Baidu's sponsored search retrieval system, which yields a remarkable revenue growth.


\section{Related Work}
Patterns are widely used in the task-oriented dialogue systems \cite{chen2017survey} \cite{li2017end}\cite{bordes2016learning}.
Given an utterance, the natural language understanding  module of the dialogue system would
parse it into semantic entity types (known as slots) and determines
the user intents. Usually, slots and intents are pre-defined in a system.
Patterns have also been studied in the paraphrase  generation and identification tasks \cite{lin2001discovery} \cite{ravichandran2002learning} \cite{pang2003syntax} \cite{szpektor2004scaling}.
\cite{pang2003syntax} describes a syntax-based algorithm that automatically builds paraphrase patterns from semantically equivalent translation sets.
\cite{zhao2008pivot} proposed a pivot approach for extracting paraphrase patterns from bilingual parallel corpora.


Knowledge Graph is commonly used in information retrieve systems \cite{wise2020covid}\cite{liu2018entity}\cite{eder2012knowledge}
and recommendation systems \cite{wang2019kgat}\cite{sun2018recurrent}\cite{wang2019multi}\cite{cao2019unifying}.
Microsoft built its Concept Graph named Probase \cite{wu2012probase} to optimize the web search and document understanding services.
\cite{mustafa2008ontology}  used the Resource Description Framework triples data to compute the thematic similarity for information retrieval.
Knowledge graph of Wikidata \cite{haase2017alexa} was applied to Amazon's
virtual assistant Alexa to offer better answers to factual questions.
A large-scale cognitive concept net was constructed by \cite{luo2020alicoco} to better define user needs and offer more intelligent shopping experience in their e-commerce platform. \cite{liu2019user} utilized query log and search click graph to discover user-centered concepts at more precise granularity that can represent users interests.


\section{Method}
Our method consists of three parts: data conceptualization,  matching via conceptual patterns, and concept-augmented discrimination.

\subsection{Data Conceptualization}
In this step, the original query/keyword will be transformed into a conceptual pattern form based on concept tagging,
which means assigning each token in the sentence with a domain concept label.
Different from the traditional named entity recognition (NER) task,
which focuses on recognizing the entities in a small number of categories such
as organizations and locations,
our concept tagging task covers entities and function words from all domains.
The taxonomy is constructed based on Baidu's Knowledge Graph, which covers 5 billion entities and 550 billion facts.
To expand the conceptual pattern's application scope, we prefer coarse concepts over refined concepts.
For example, the entity \emph{double eyelid surgery} belongs to a refined
concept of \emph{[eye plastic]}, which
has a coarse hypernym concept named \emph{[aesthetic surgery]}. Considering that most of the \emph{eye plastic} related paraphrase patterns can also be applied in \emph{aesthetic surgery} scenarios,
the coarse concept of \emph{[aesthetic surgery]} is
selected.


\begin{figure}[!h]
  \centering
  \resizebox{1.2\totalheight}{!}{
      \includegraphics[width=0.56\textwidth]{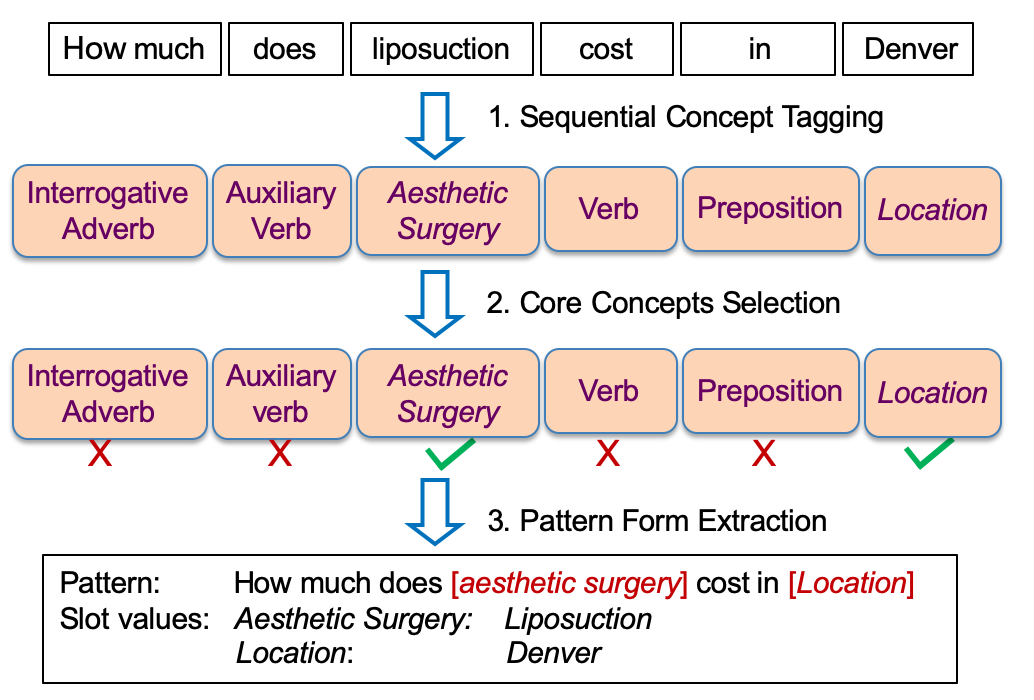}

   }
   \caption{The process of commercially customized concept tagging.}
  \label{fig:ecom_labeling}
    \vspace{-3mm}
\end{figure}

As is shown in Figure \ref{fig:ecom_labeling}, the concept tagging procedure consists of three steps:
\begin{itemize}
  \item Firstly, each token in the sentence is assigned with a concept label based on concept tagging.
  \item Secondly, core concepts are selected for further generalization and the remaining concepts are neglected.
The core concept set is a manually selected subset of the ontology, which focuses on commercial entity related categories such as education, traveling, cosmetology, food, and so on, while function words like adverb and preposition are overlooked.
  \item Finally, the original sentence is extracted into a pattern form mixed with concept slots and normal texts,
and slots' corresponding entities are represented as slot values.
\end{itemize}

Therefore, the query in the figure \emph{how much does liposuction cost in Denver}
will be conceptualized as
\emph{how much does [aesthetic surgery] cost in [location]} with slot values \emph{[aesthetic surgery: liposuction] and [location: Denver]}.
\begin{figure*}[t]
  \centering
  \resizebox{1.0\totalheight}{!}{
      \includegraphics[width=1.6\textwidth]{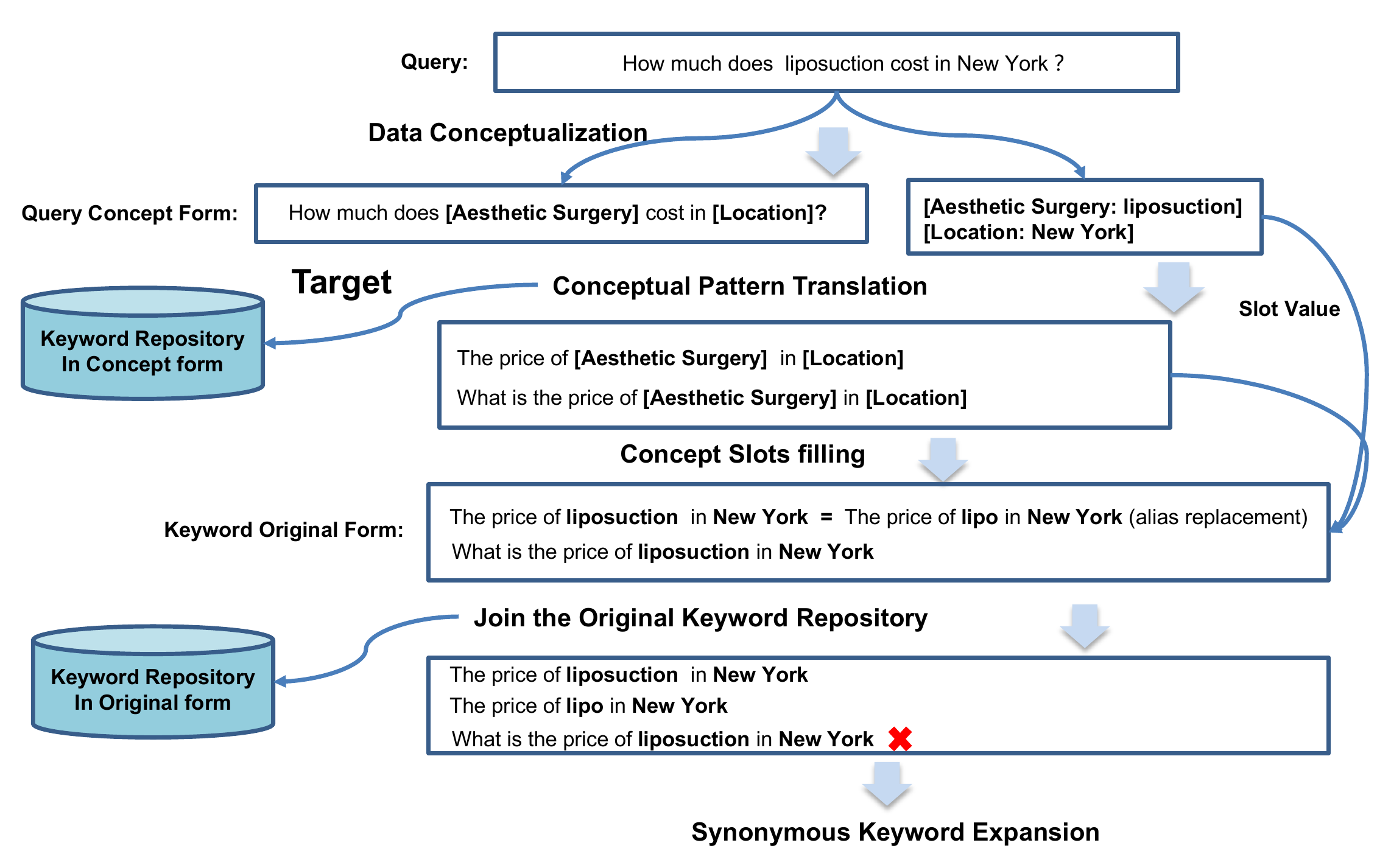}
       }
   \caption{The specific query-\emph{keyword} matching procedures.}
  \label{fig:main_idea}
    \vspace{-3mm}
\end{figure*}

\subsection{Matching via Conceptual Patterns}
The basic idea of our framework is to transform queries and \emph{keywords} into conceptual patterns
and conduct the synonymous query-\emph{keyword} matching through conceptual pattern matching.
\subsubsection{Pattern-to-Pattern Model Training}

The matching between query pattern and \emph{keyword} pattern is based on a pattern-to-pattern
translation model. Given a high quality paraphrase data set,
data conceptualization is performed on both sides to obtain parallel pattern training data.
For example, the original instance
  (\emph{how much does double eyelid surgery cost in Los Angeles
  = the price of double-fold eyelid operation in Los Angeles}) is changed into
(\emph{the price of [Aesthetic Surgery] in [Location]}
= \emph{how much does [Aesthetic Surgery] cost in [Location]}).
For high precision concerns, we further clean the conceptualized parallel pattern data
with the requirement that the left pattern should be strictly aligned with the right pattern,
not only the number of concept slots but also the corresponding entities should be the same.


And then a transformer-based translation model is trained on the conceptualized parallel data.
In detail, concept slots are considered as normal tokens and added to the model's vocabulary dictionary.
The structure of the model follows the common sequence to sequence framework \cite{Sutskever2014}, where
the encoder encodes the source sentence into a list of hidden states and
the decoder generates word one by one until a special end symbol is generated.

\subsubsection{Pattern-Based Keyword Matching}

When an ad-hoc query comes, the following
four steps are carried out to generate its synonymous candidates.

The first step is query conceptualization. It follows the same procedure as
data conceptualization mentioned above.

The second step is pattern matching. The pattern translation model is utilized
to link the query's pattern with the \emph{keyword} pattern repository, which is
a conceptualized version of the original \emph{keyword} repository.
A prefix-tree-based targeted decoding trick is utilized \cite{lian2019end} to ensure
that all the decoded hypotheses are valid \emph{keyword} patterns.

The third step is \emph{keyword} pattern instantiation.
The concept slots in the previously retrieved \emph{keyword} patterns
are replaced with the query's corresponding entities. To increase recall,
the entities' aliases from the knowledge database are also utilized.
Since the instantiated pattern might not
be real \emph{keywords}, we further join them with the original \emph{keyword} repository to remove illegal results.

The final step is synonymous \emph{keyword} expansion.
Considering that there is also a number of synonymous
relationships which could not be strictly expressed as aligned pattern forms,
a synonymous expanding process is performed in the end.
In detail, synonymous keywords are clustered in advance by utilizing \emph{keyword} to \emph{keyword}
synonymous retrieval method \cite{lian2020para}.
If a previously retrieved \emph{keyword} belongs to a
\emph{keyword} cluster, the whole \emph{keyword} cluster will be merged into the final candidate queue.

The whole matching process is illustrated in Figure \ref{fig:main_idea}.
When an ad-hoc query \emph{How much does liposuction cost in New York} arrives,
it is firstly conceptualized as \emph{How much does [Aesthetic Surgery] cost in [Location]} with slot values \emph{[Aesthetic Surgery: liposuction] and [Location: Denver]}, entities' aliases such as liposuction(lipo)
are fetched from the knowledge database in the meanwhile.
Then a pattern translation model is utilized to link the query's pattern with
the conceptual \emph{keyword} repository, the following two synonymous \emph{keyword} patterns are produced:
\emph{The price of [Aesthetic Surgery] in [Location]} and \emph{What is the price of [Aesthetic Surgery] in [Location]}.
In the next step, concept slots in these \emph{keyword} patterns
are replaced with the corresponding query entities and their aliases, which generates
three sentences: \emph{The price of liposuction in New York} , \emph{The price of lipo in New York} and \emph{What is the price of liposuction in New York}.
Then, we join these sentences with the real \emph{keyword} repository to remove invalid keywords, and \emph{What is the price of liposuction in New York}
is removed in the example.
Finally, synonymous keywords expansion is performed.

\subsection{Concept-Augmented Discrimination}
\label{sec:data_aug}
To guarantee the final query-\emph{keyword} pairs' synonymous quality, an end2end
concept augmented discrimination is performed based on a domain fine-tuned BERT \cite{devlin2018bert} model.

Since the discriminant model also faces the long-tail instance generalization problem,
directly using it might misjudge lots of conceptually retrieved cases.
Therefore, we augment the fine-tuning data by replacing the entities with same-concept entities.
In detail, for the original synonymous query-\emph{keyword} instances,
we replace the aligned concept slot of query and \emph{keyword} with the same rare entity to get augmented positive cases
and we replace them with different rare entities to get augmented negative cases,
where literally confusable entities are particularly used.
For the original negative query-\emph{keyword} instances, with a probability of 50\%, we replace
the aligned concept slots with the same rare entity, and with a probability of 50\%, we replace
them with different rare entities.
Offline experiments show that this method greatly improves the model's robustness for
long-tail instances.

\section{Offline Experiment}

\subsection{Matching via Conceptual Patterns}
\subsubsection{Dataset for The Translation Model}
Data conceptualization is performed on one day's query-keyword {exact matching}  weblog,
and 22 million paraphrasing pattern pairs $\widetilde{D}^{gen}$ are sampled from it.
This pattern pair dataset is split into two parts, $\widetilde{D}^{gen}_{train}$ for training
and $\widetilde{D}^{gen}_{dev}$ for development.
And the corresponding original datasets are denoted as $D^{gen}_{train}$ and $D^{gen}_{dev}$.
The baseline and the conceptual translation models $M^{gen}$ and $\widetilde{M}^{gen}$ are separately trained on $D^{gen}_{train}$ and
$\widetilde{D}^{gen}_{train}$.
To evaluate the translation model's generalization ability on the
entity-level
long-tail cases, we construct a query test set $D^{gen}_{test}$ by joining all the queries
in $D^{gen}_{train}$ with one week's query weblog based on their contained entities.
For simplicity,
we only consider queries having one entity. Queries in
$D^{gen}_{test}$ are grouped into 4 buckets according to their entity frequencies in $D^{gen}_{train}$.
Denote the frequency as $f$,  four frequency intervals are specially considered: $1 < f \leq 10, 10 < f \leq 100, 100 < f \leq 1000, f > 1000$.  And the queries located in the first bucket are considered
as the long-tail query set $D^{gen}_{test0}$.

\subsubsection{Implementation Details}
The translation model's word embeddings are randomly initialized. The vocabulary contains 200,000 frequent words and 284 concept slots.
The word embedding dimension and the number of hidden units are set to 512. The max sequence length is limited to 12 at the word level.
Both the encoder and decoder are  implemented with transformers having 4
layers and 8 heads.
The model’s cross-entropy loss is minimized by Adam \cite{kingma2014adam} with initial learning rate
of $5 \times 10^{-5}$ and the batch size is 64.

\subsubsection{Results}
Setting beam size to 50, $M^{gen}$ and $\widetilde{M}^{gen}$ are utilized to decode towards queries in $D^{gen}_{test}$.
For $\widetilde{M}^{gen}$, the concept slots in the decoded patterns
are replaced with the original slot values in the query.
For each bucket in $D^{gen}_{test}$,
we sample 500 generated cases for human synonymous binary evaluation.
As is shown in the
Table \ref{table:result_translation_model},
$\widetilde{M}^{gen}$ performs much better than
the $M^{gen}$, and it outperforms  $M^{gen}$ by 26 absolute points on the long-tail
query dataset $D^{gen}_{test0}$. We can also see that
the raw translation model's performance
is significantly influenced by the entity frequency in the training dataset.
In contrast, $\widetilde{M}^{gen}$ 's performance is insensitive to
the entity frequency.

\begin{table}[h]
\vskip 2mm
\centering
\begin{tabular}{cccccccccc}
\hline
\textbf{Frequency}  & $M^{gen}$   & $\widetilde{M}^{gen}$  \\ \hline
1-10&37.4\%& 63.4\% \\
10-100& 53\%& 63.8\% \\
100-1000& 65.6\%& 66.4\% \\
1000 or more& 68.8\%& 68.6\% \\
 \hline
\\
\end{tabular}
\caption{Accuracy performances of the two translation models on $D^{gen}_{test}$.
  The frequency
denotes the query's entity occurrence frequencies in $D^{gen}_{train}$. }
\label{table:result_translation_model}
\centering
\end{table}

\subsection{Concept-Augmented Discrimination}
\subsubsection{Dataset for The Discriminant Model}
460,000 query-keyword pairs $D^{dis}$ are sampled from the sponsored matching weblog for human
synonymous evaluation,
which covers all three matching types: \emph{exact match}, \emph{phrase match}, and \emph{broad match} with a  proportion ratio of 2:1:1. $D^{dis}$ is further split
into three parts: $D^{dis}_{train}$ for training,  $D^{dis}_{dev}$ for developing and $D^{dis}_{test}$ for testing. Then concept augmentation procedure mentioned in subsection \ref{sec:data_aug} is performed on $D^{dis}_{train}$.
55,200 augmented cases are sampled, which accounts for 12\% of $D^{dis}_{train}$,
and merged with $D^{dis}_{train}$ to get $\hat{D}^{dis}_{train}$.
$D^{dis}_{test}$ can be considered as a test set from a global perspective, however,
since we care more about the discriminant model's  generalization ability on the conceptually retrieved long-tail cases,
another test data $\bar{D}^{dis}_{test}$ is constructed by sampling 1000 cases from  $\widetilde{M}^{gen}$'s generated cases on $D^{gen}_{test0}$
for human evaluation.

\subsubsection{Implementation Details}
The paraphrase discriminant model is a binary classifier implemented with BERT. It takes a query-keyword pair separated by a special token as input and predicts 1 if the pair is synonymous and 0 if not. The model contains 24 layers, 16 self-attention heads and the hidden dimension size is 1024. The parameters
are initialized with ERNIE \cite{ernie_baidu}, a well-known Chinese
pre-trained transformer published by Baidu.
The baseline model $M^{dis}$ and the concept-augmented model $\hat{M}^{dis}$ are
separately fine-tuned on $D_{train}^{dis}$ and $\hat{D}^{dis}_{train}$.
The fine-tuned loss is minimized by Adam with initial learning rate of $5
\times 10^{-6}$ and the batch size is 128.


\subsubsection{Results}

We focus on two indicators: AUC (area under the curve) and the recall under 95\%/70\% precision.
Table \ref{table:result_dis_model_normal_test} shows that the concept-augmented model outperforms
the baseline model with 16.24 points for the recall under 70\% precision on $\bar{D}^{dis}_{test}$.
Since the data distribution between $\bar{D}^{dis}_{test}$ and $D_{test}^{dis}$ differs a lot,
where long-tail cases
occupy a small proportion in the general test dataset $D_{test}^{dis}$,
the indicators on the $D_{test}^{dis}$ have just been slighted improved.

\begin{table}[H]
\vskip -1mm
\centering
\begin{tabular}{cccccccccc}
\hline
\textbf{Model}  & \textbf{AUC-G}   & \textbf{Recall-G} & \textbf{AUC-L} & \textbf{Recall-L}\\ \hline
 $M^{dis}$&  97.53\%& 75.86\% & 59.85\% & 61.21\% \\
 $\hat{M}^{dis}$& 97.53\% & 76.73\% & 65.31\% & 77.45\%  \\ \hline
\\
\end{tabular}
\caption{Results of different discriminant models on $D_{test}^{dis}$ and
    $\bar{D}_{test}^{dis}$. AUC-G and Recall-G denote the AUC value and recall
    ratio under the precision of 95\% on the global test set $D_{test}^{dis}$.
    AUC-L and Recall-L denote the AUC value and recall ratio under the precision of 70\% on the long-tail test set $\bar{D}_{test}^{dis}$.}
\label{table:result_dis_model_normal_test}
\centering
\vskip -6mm
\end{table}

Table \ref{table:result_different_num_augmented_data} shows the performances of the discriminant model
$\hat{M}^{dis}$ with different proportion of augmented data.
We can see that the model's recall ratio on $\bar{D}_{test}^{dis}$ grows as the proportion increases.
However, 
too much augmented data would destroy the original data's distribution,
which results in a bad performance on the global test set $D_{test}^{dis}$.
As we can see in the table, the performance on $D_{test}^{dis}$ peaks around the proportion at 12\%.

\begin{table}[H]
\vskip 1mm
\centering
\begin{tabular}{cccccccccc}
\hline
\textbf{Proportion}  & \textbf{Recall-G}   & \textbf{Recall-L} \\ \hline
8\% & 76.14\% & 72.18\%    \\
10\% & 76.58\% & 76.95\%    \\
12\% & \textbf{76.73\%} & 77.45\%    \\
16\% & 76.17\% & \textbf{81.72\%}  \\ \hline
\\
\end{tabular}
\caption{The performances of the discriminant model $\hat{M}^{dis}$ with different proportion of the augmentation data.}
\label{table:result_different_num_augmented_data}
\centering
\vskip -6mm
\end{table}

\begin{table*}[h]
\vskip 4mm
\centering
\begin{tabular}{ccccc}
\hline
\textbf{Query} & \textbf{Pattern}  & \textbf{Entity Freq} \\ \hline
\underline{海马丘比特}质量怎么样& [品牌]质量怎么样 & 3   \\
(How about the quality of the \underline{Haima Qiubite})& (How about the quality of [Brand])& \\

\underline{欧米伽3}哪里有 & [食物营养素]哪里有& 2 \\
(Where can we buy \underline{OMEGA-3}) & (Where can we buy [Food Nutrients])& \\

\underline{炮附片}的功效& [植物类中药材]的功效& 1 \\
(The effect of \underline{Paofu tablets}) & (The effect of [Plant Chinese Medicinal Materials]) & \\

\underline{哈蜜瓜}孕妇能吃么 & [饮食]孕妇能吃么& 1  \\
(Can pregnant women eat \underline{cantaloupe}) & (Can pregnant women eat [Food]) & \\

{伦敦糕}的做法与配方& [饮食]的做法与配方 & 1 \\
(The practice and recipe of \underline{London cake}) & (The practice and recipe of [Food]) & \\ \hline
\\
\end{tabular}
\caption{Some typical long-tail queries and their conceptual patterns, where entities are underlined and concepts are enclosed in square brackets. Besides, Entity Freq denotes the entity's frequency in $D^{gen}_{train}$.}
\label{table:case_study_trans_model}
\centering
\vskip -6mm
\end{table*}

\begin{table*}[h]
\vskip 4mm
\centering
\begin{tabular}{ccccc}
\hline
\textbf{Query} &\textbf{$M^{gen}$} & \textbf{$\widetilde{M}^{gen}$} \\ \hline
海马丘比特质量怎么样& 海马好吗 & 海马丘比特这车怎么样 \\
(How about the quality of the Haima Qiubite)& (How is Haima)& (How about the car named Haima Qiubite) \\
& 海马下垂怎么样& 海马丘比特质量好不好\\
& (How about hippocampus sagging)& (Is the quality of Haima Qiubite good)\\
& 海马法语怎么样& 海马丘比特质量如何\\
& (How about Haima French)& (How is the quality of Haima Qiubite)\\
哈蜜瓜孕妇能吃么 & 孕妇能吃橙子吗 & 怀孕了能吃哈蜜瓜吗 \\
(Can pregnant women eat cantaloupe) & (Can pregnant women eat oranges)& (Can I eat cantaloupe while pregnant)\\
& 孕妇可以吃益母草吗& 孕妇可以吃哈蜜瓜吗\\
& (Can pregnant women eat leonotis)& (Can pregnant women eat cantaloupe)\\
& 孕妇能吃黄芪吗& 怀孕期间可以吃哈蜜瓜吗\\
& (Can pregnant women eat astragalus)& (Can I eat cantaloupe during pregnancy)\\ \hline
\\
\end{tabular}
\caption{The translation results of typical cases.}
\label{table:typical_translation_cases}
\centering
\vskip -6mm
\end{table*}

\begin{table*}[h]
\vskip 4mm
\centering
\begin{tabular}{ccccc}
\hline
\textbf{Query} & \textbf{Bidword}  & \textbf{$M^{dis}$} & \textbf{$\hat{M}^{dis}$} & \textbf{Label} \\ \hline
血HCG检查多少钱& 血疮检查费用 & 0.6 & 0.03 & 0   \\
(How much is the blood HCG test)& (Blood sore inspection fees)&  & &    \\
花牛苹果什么科 & 嘎拉苹果属于什么科& 0.67 & 0.07 & 0 \\
(What family is Huaniu Apple) & (What family does Gala Apple belong to)& &  &  \\
道奇汽车是哪国生产 & 道奇汽车产地在哪里& 0.75 & 0.95 & 1 \\
(Where are Dodge cars made) & (Where is the origin of Dodge cars) & &  &  \\ \hline
\\
\end{tabular}
\caption{The prediction scores of different discriminant models on some typical long-tail query-\emph{keyword} cases. The label 1 denotes that the query-keyword pair is synonymous, while 0 stands for the opposite.  }
\label{table:case_study_dis_model}
\centering
\vskip -3mm
\end{table*}

\subsection{Case Study and Discussion}
\subsubsection{Case Study}
Table \ref{table:case_study_trans_model} shows some typical long-tail queries
and their corresponding conceptual patterns.
Table \ref{table:typical_translation_cases} further shows the
top 3 results decoded by $M^{gen}$ and $\tilde{M}^{gen}$.
We can see that the traditional translation model performs badly on the
long-tail queries, and it can not preserve the source entities in the decoding phase.
For example, in the case of \emph{Can pregnant women eat cantaloupe}, $M^{gen}$'s decoding
results have changed the entity
\emph{cantaloupe} into \emph{oranges} or other nonsynonymous items. However, the concept-based decoding precisely
retains the entity.

Table \ref{table:case_study_dis_model} shows the discriminant model's inference scores on some typical long-tail cases.
We can see that the model's predicted scores have been well adjusted, where the new model
predicts much higher scores for positive cases and much lower scores for
negative ones after fine-tuning on the conceptually augmented data.

\subsubsection{Pattern Coverage}
The synonymous pattern's coverage plays a crucial role in the practical application of our method.
Our human evaluation statistics shows that nearly 70\% of the
existing commercial paraphrases can be transformed into conceptual forms.
As a matter of fact, people's commercial intentions are relatively concentrated in the sponsored search scenario,
and most of these paraphrases can be expressed as steady pattern forms.
For example, goods' price is one of the most common commercial intentions,
which can be easily abstracted as \emph{how much is [Goods]}.
Table \ref{table:case_study_trans_model} shows some typical commercial patterns
related to goods' quality, food recipe, and medicine functions.

\begin{figure*}[h]
  \centering
  \resizebox{1.0\totalheight}{!}{
      \includegraphics[width=1.6\textwidth]{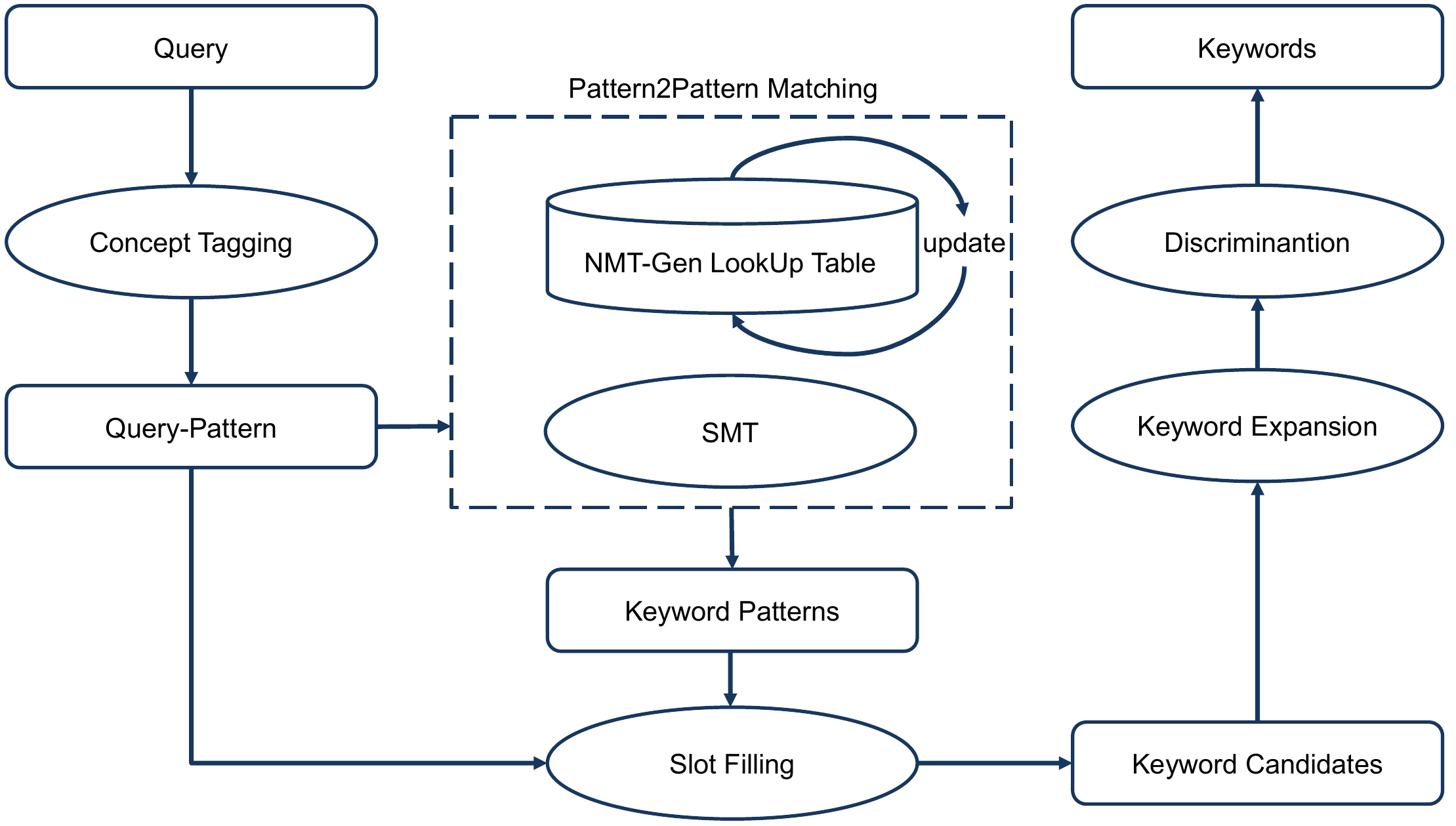}
       }
   \caption{The online implementation framework of the concept knowledge-driven retrieval module.}
  \label{fig:KITR}
    \vspace{-3mm}
\end{figure*}

\section{Online Experiment}
As is illustrated in Figure \ref{fig:KITR},
we deployed a new module named CKBR(Concept Knowledge Based Retrieval)
on Baidu's sponsored search engine for retrieving \emph{exact-match-type} keywords.

For latency concerns, the pattern matching phase is implemented under an online-offline mixed architecture.
For frequent query patterns, their matched \emph{keyword} patterns are computed ahead of time with a complex transformer model,
and these results are saved in a lookup table for fast online retrieval.
And this offline process is repeated periodically to follow the changes in the query population and \emph{keyword} supply over time.
For infrequent queries, a phrase-based statistical machine translation(PBSMT) framework is utilized to generate \emph{keyword} patterns from scratch.
Although neural machine translation(NMT) has outperformed PBSMT in lots of translation tasks, we find that PBSMT is still a
cost-effective choice for the industry. In contrast with NMT, it is much faster and no GPU is required. As a matter of fact,
our PBSMT is able to decode out 1200 synonymous variations in less than 60ms. Nevertheless, a simple two-layer GRU \cite{chung2014empirical} gated NMT can just
decode out 30 synonymous variations in the same time window(The performance test is conducted on
a machine equipped with a 12-core Intel(R) Xecon(R) E5-2620 v3 clocked at 2.40GHz,
a RAM of 128G, and 16 Tesla K40m.  PBSMT runs with 10 CPU threads and the stack
size is 100. NMT runs
in a single GPU with a beam size of 30). Another advantage of SMT is that
all of the alignments are stored in an editable and readable phrase table, which makes it explainable and controllable.

The online discriminant model is implemented with a transformer having 2 layers and 4 heads, with a hidden size of 128.
This model is distilled from the previous concept-augmented discriminant model $\hat{M}^{dis}$.
Considering the chapter space, we would not elaborate on the details.

\begin{table}[h]
\vskip 2mm
\centering
\begin{tabular}{cccccccccc}
\hline
\textbf{Platform}  & \textbf{SHOW}   & \textbf{CPM}  & \textbf{CTR} &
    \textbf{SHOW-EXACT}  \\ \hline
Mobile&  0.54\%& 1.51\%& 0.92\% &9.14\%       \\
Desktop& 1.58\% & 2.52\%& 1.43\%  &8.42\%      \\ \hline
\\
\end{tabular}
\caption{The online A/B test results of the CKBR module, which indicate the relative improvements over the current system.}
\label{online_ab_exp_result}
\centering
\vskip -5.4mm
\end{table}

A 10 days' real online A/B test is deployed on two platforms of Baidu's sponsored search engines, which corresponds
to query flows from desktops and mobile devices. We focus on the following metrics.

\begin{itemize}
\item SHOW denotes the total ads' number shown to users
\item CTR = $\frac{\rm{CLICK}}{\rm{SEARCH}}$ denotes the average click-through rate. CLICK denotes the number of all ads' clicks and SEARCH denotes the
  number of all searches.
\item CPM = $\frac{\rm{REVENUE}}{\rm{SEARCH}} \times 1000$, which denotes average revenue per one thousand searches received by the search engine.
\item SHOW-EXACT denotes the number of shown ads under the \emph{exact match type}.
\end{itemize}

As is shown in the Table \ref{online_ab_exp_result}, the CKBR module leads
to an evident improvement of SHOW-EXACT by 8.42\% on desktop flows and 9.14\% on mobile flows.
Besides, it also yields a significant growth in CPM with 2.52\% on desktop flows and 1.51\% on mobile flows.
In the meanwhile, a quality evaluation has been conducted, where
600 query-\emph{keyword} cases under the \emph{exact match type}
are sampled from the system's weblog and are sent for synonymous binary judgment. And the evaluation result shows that
the synonym accuracy has increased by 0.7\%.

\section{Conclusions}
In this paper, we have developed a novel knowledge-driven framework for addressing the
synonymous keywords retrieval problem
in sponsored search.
Under this framework, the synonymous transformation can be understood as a combination of pattern transformation and alias replacement.
Based on a large Chinese knowledge graph, we are able to conceptualize most of the commercial
query-\emph{keyword} synonymous pairs into abstract patterns.
A deep translation model is trained on this conceptualized data to capture the synonymous pattern variations.
Our offline experiments show that the new framework performs much
better on the long-tail cases.
The whole framework has been implemented in Baidu's keywords retrieval system based on a NMT/SMT mixed architecture,
and a significant improvement in revenue has been yielded without degrading the quality of
users' experience.
This method's application scope is not limited in the synonymous retrieval under \emph{exact match},
exploration in \emph{phrase match} and \emph{broad match} is in progress.
We hope our method would shed some light on the further design of the industrial sponsored search system.
\bibliographystyle{ACM-Reference-Format}
\bibliography{pattern_retrieval.bib}


\begin{thebibliography}{29}


\ifx \showCODEN    \undefined \def \showCODEN     #1{\unskip}     \fi
\ifx \showDOI      \undefined \def \showDOI       #1{#1}\fi
\ifx \showISBNx    \undefined \def \showISBNx     #1{\unskip}     \fi
\ifx \showISBNxiii \undefined \def \showISBNxiii  #1{\unskip}     \fi
\ifx \showISSN     \undefined \def \showISSN      #1{\unskip}     \fi
\ifx \showLCCN     \undefined \def \showLCCN      #1{\unskip}     \fi
\ifx \shownote     \undefined \def \shownote      #1{#1}          \fi
\ifx \showarticletitle \undefined \def \showarticletitle #1{#1}   \fi
\ifx \showURL      \undefined \def \showURL       {\relax}        \fi
\providecommand\bibfield[2]{#2}
\providecommand\bibinfo[2]{#2}
\providecommand\natexlab[1]{#1}
\providecommand\showeprint[2][]{arXiv:#2}

\bibitem[\protect\citeauthoryear{Bordes, Boureau, and Weston}{Bordes
  et~al\mbox{.}}{2016}]%
        {bordes2016learning}
\bibfield{author}{\bibinfo{person}{Antoine Bordes}, \bibinfo{person}{Y-Lan
  Boureau}, {and} \bibinfo{person}{Jason Weston}.}
  \bibinfo{year}{2016}\natexlab{}.
\newblock \showarticletitle{Learning end-to-end goal-oriented dialog}.
\newblock \bibinfo{journal}{\emph{arXiv preprint arXiv:1605.07683}}
  (\bibinfo{year}{2016}).
\newblock


\bibitem[\protect\citeauthoryear{Cao, Wang, He, Hu, and Chua}{Cao
  et~al\mbox{.}}{2019}]%
        {cao2019unifying}
\bibfield{author}{\bibinfo{person}{Yixin Cao}, \bibinfo{person}{Xiang Wang},
  \bibinfo{person}{Xiangnan He}, \bibinfo{person}{Zikun Hu}, {and}
  \bibinfo{person}{Tat-Seng Chua}.} \bibinfo{year}{2019}\natexlab{}.
\newblock \showarticletitle{Unifying knowledge graph learning and
  recommendation: Towards a better understanding of user preferences}. In
  \bibinfo{booktitle}{\emph{The world wide web conference}}.
  \bibinfo{pages}{151--161}.
\newblock


\bibitem[\protect\citeauthoryear{Chen, Liu, Yin, and Tang}{Chen
  et~al\mbox{.}}{2017}]%
        {chen2017survey}
\bibfield{author}{\bibinfo{person}{Hongshen Chen}, \bibinfo{person}{Xiaorui
  Liu}, \bibinfo{person}{Dawei Yin}, {and} \bibinfo{person}{Jiliang Tang}.}
  \bibinfo{year}{2017}\natexlab{}.
\newblock \showarticletitle{A survey on dialogue systems: Recent advances and
  new frontiers}.
\newblock \bibinfo{journal}{\emph{Acm Sigkdd Explorations Newsletter}}
  \bibinfo{volume}{19}, \bibinfo{number}{2} (\bibinfo{year}{2017}),
  \bibinfo{pages}{25--35}.
\newblock


\bibitem[\protect\citeauthoryear{Chung, Gulcehre, Cho, and Bengio}{Chung
  et~al\mbox{.}}{2014}]%
        {chung2014empirical}
\bibfield{author}{\bibinfo{person}{Junyoung Chung}, \bibinfo{person}{Caglar
  Gulcehre}, \bibinfo{person}{KyungHyun Cho}, {and} \bibinfo{person}{Yoshua
  Bengio}.} \bibinfo{year}{2014}\natexlab{}.
\newblock \showarticletitle{Empirical evaluation of gated recurrent neural
  networks on sequence modeling}.
\newblock \bibinfo{journal}{\emph{arXiv preprint arXiv:1412.3555}}
  (\bibinfo{year}{2014}).
\newblock


\bibitem[\protect\citeauthoryear{Devlin, Chang, Lee, and Toutanova}{Devlin
  et~al\mbox{.}}{2018}]%
        {devlin2018bert}
\bibfield{author}{\bibinfo{person}{Jacob Devlin}, \bibinfo{person}{Ming-Wei
  Chang}, \bibinfo{person}{Kenton Lee}, {and} \bibinfo{person}{Kristina
  Toutanova}.} \bibinfo{year}{2018}\natexlab{}.
\newblock \showarticletitle{Bert: Pre-training of deep bidirectional
  transformers for language understanding}.
\newblock \bibinfo{journal}{\emph{arXiv preprint arXiv:1810.04805}}
  (\bibinfo{year}{2018}).
\newblock


\bibitem[\protect\citeauthoryear{Eder}{Eder}{2012}]%
        {eder2012knowledge}
\bibfield{author}{\bibinfo{person}{Jeffrey~Scott Eder}.}
  \bibinfo{year}{2012}\natexlab{}.
\newblock \bibinfo{title}{Knowledge graph based search system}.
\newblock
\newblock
\newblock
\shownote{US Patent App. 13/404,109.}


\bibitem[\protect\citeauthoryear{Guo, Xu, Cheng, and Li}{Guo
  et~al\mbox{.}}{2009}]%
        {guo2009named}
\bibfield{author}{\bibinfo{person}{Jiafeng Guo}, \bibinfo{person}{Gu Xu},
  \bibinfo{person}{Xueqi Cheng}, {and} \bibinfo{person}{Hang Li}.}
  \bibinfo{year}{2009}\natexlab{}.
\newblock \showarticletitle{Named entity recognition in query}. In
  \bibinfo{booktitle}{\emph{Proceedings of the 32nd international ACM SIGIR
  conference on Research and development in information retrieval}}.
  \bibinfo{pages}{267--274}.
\newblock


\bibitem[\protect\citeauthoryear{Haase, Nikolov, Trame, Kozlov, and
  Herzig}{Haase et~al\mbox{.}}{2017}]%
        {haase2017alexa}
\bibfield{author}{\bibinfo{person}{Peter Haase}, \bibinfo{person}{Andriy
  Nikolov}, \bibinfo{person}{Johannes Trame}, \bibinfo{person}{Artem Kozlov},
  {and} \bibinfo{person}{Daniel~M Herzig}.} \bibinfo{year}{2017}\natexlab{}.
\newblock \showarticletitle{Alexa, Ask Wikidata! Voice Interaction with
  Knowledge Graphs using Amazon Alexa.}. In
  \bibinfo{booktitle}{\emph{International Semantic Web Conference (Posters,
  Demos \& Industry Tracks)}}.
\newblock


\bibitem[\protect\citeauthoryear{Hendrycks, Liu, Wallace, Dziedzic, Krishnan,
  and Song}{Hendrycks et~al\mbox{.}}{2020}]%
        {hendrycks2020pretrained}
\bibfield{author}{\bibinfo{person}{Dan Hendrycks}, \bibinfo{person}{Xiaoyuan
  Liu}, \bibinfo{person}{Eric Wallace}, \bibinfo{person}{Adam Dziedzic},
  \bibinfo{person}{Rishabh Krishnan}, {and} \bibinfo{person}{Dawn Song}.}
  \bibinfo{year}{2020}\natexlab{}.
\newblock \showarticletitle{Pretrained transformers improve out-of-distribution
  robustness}.
\newblock \bibinfo{journal}{\emph{arXiv preprint arXiv:2004.06100}}
  (\bibinfo{year}{2020}).
\newblock


\bibitem[\protect\citeauthoryear{Kingma and Ba}{Kingma and Ba}{2015}]%
        {kingma2014adam}
\bibfield{author}{\bibinfo{person}{Diederik~P Kingma} {and}
  \bibinfo{person}{Jimmy Ba}.} \bibinfo{year}{2015}\natexlab{}.
\newblock \showarticletitle{Adam: A method for stochastic optimization}.
\newblock \bibinfo{journal}{\emph{International Conference on Learning
  Representations (ICLR)}} (\bibinfo{year}{2015}).
\newblock


\bibitem[\protect\citeauthoryear{Li, Chen, Li, Gao, and Celikyilmaz}{Li
  et~al\mbox{.}}{2017}]%
        {li2017end}
\bibfield{author}{\bibinfo{person}{Xiujun Li}, \bibinfo{person}{Yun-Nung Chen},
  \bibinfo{person}{Lihong Li}, \bibinfo{person}{Jianfeng Gao}, {and}
  \bibinfo{person}{Asli Celikyilmaz}.} \bibinfo{year}{2017}\natexlab{}.
\newblock \showarticletitle{End-to-end task-completion neural dialogue
  systems}.
\newblock \bibinfo{journal}{\emph{arXiv preprint arXiv:1703.01008}}
  (\bibinfo{year}{2017}).
\newblock


\bibitem[\protect\citeauthoryear{Lian, Chen, Hu, Zhang, Yan, Tong, Han, Guan,
  Li, Cao, et~al\mbox{.}}{Lian et~al\mbox{.}}{2019}]%
        {lian2019end}
\bibfield{author}{\bibinfo{person}{Yijiang Lian}, \bibinfo{person}{Zhijie
  Chen}, \bibinfo{person}{Jinlong Hu}, \bibinfo{person}{Kefeng Zhang},
  \bibinfo{person}{Chunwei Yan}, \bibinfo{person}{Muchenxuan Tong},
  \bibinfo{person}{Wenying Han}, \bibinfo{person}{Hanju Guan},
  \bibinfo{person}{Ying Li}, \bibinfo{person}{Ying Cao}, {et~al\mbox{.}}}
  \bibinfo{year}{2019}\natexlab{}.
\newblock \showarticletitle{An end-to-end Generative Retrieval Method for
  Sponsored Search Engine--Decoding Efficiently into a Closed Target Domain}.
\newblock \bibinfo{journal}{\emph{arXiv preprint arXiv:1902.00592}}
  (\bibinfo{year}{2019}).
\newblock


\bibitem[\protect\citeauthoryear{Lian, You, Wu, Liu, and Jia}{Lian
  et~al\mbox{.}}{2020}]%
        {lian2020para}
\bibfield{author}{\bibinfo{person}{Yijiang Lian}, \bibinfo{person}{Zhenjun
  You}, \bibinfo{person}{Fan Wu}, \bibinfo{person}{Wenqiang Liu}, {and}
  \bibinfo{person}{Jing Jia}.} \bibinfo{year}{2020}\natexlab{}.
\newblock \showarticletitle{Retrieve Synonymous keywords for Frequent Queries
  in Sponsored Search in a Data Augmentation Way}.
\newblock \bibinfo{journal}{\emph{arXiv preprint arXiv:2008.01969}}
  (\bibinfo{year}{2020}).
\newblock


\bibitem[\protect\citeauthoryear{Lin and Pantel}{Lin and Pantel}{2001}]%
        {lin2001discovery}
\bibfield{author}{\bibinfo{person}{Dekang Lin} {and} \bibinfo{person}{Patrick
  Pantel}.} \bibinfo{year}{2001}\natexlab{}.
\newblock \showarticletitle{Discovery of inference rules for
  question-answering}.
\newblock \bibinfo{journal}{\emph{Natural Language Engineering}}
  \bibinfo{volume}{7}, \bibinfo{number}{4} (\bibinfo{year}{2001}),
  \bibinfo{pages}{343--360}.
\newblock


\bibitem[\protect\citeauthoryear{Liu, Guo, Niu, Wang, Xu, Lin, Lai, and Xu}{Liu
  et~al\mbox{.}}{2019}]%
        {liu2019user}
\bibfield{author}{\bibinfo{person}{Bang Liu}, \bibinfo{person}{Weidong Guo},
  \bibinfo{person}{Di Niu}, \bibinfo{person}{Chaoyue Wang},
  \bibinfo{person}{Shunnan Xu}, \bibinfo{person}{Jinghong Lin},
  \bibinfo{person}{Kunfeng Lai}, {and} \bibinfo{person}{Yu Xu}.}
  \bibinfo{year}{2019}\natexlab{}.
\newblock \showarticletitle{A User-Centered Concept Mining System for Query and
  Document Understanding at Tencent}. In \bibinfo{booktitle}{\emph{Proceedings
  of the 25th ACM SIGKDD International Conference on Knowledge Discovery \&
  Data Mining}}. \bibinfo{pages}{1831--1841}.
\newblock


\bibitem[\protect\citeauthoryear{Liu, Xiong, Sun, and Liu}{Liu
  et~al\mbox{.}}{2018}]%
        {liu2018entity}
\bibfield{author}{\bibinfo{person}{Zhenghao Liu}, \bibinfo{person}{Chenyan
  Xiong}, \bibinfo{person}{Maosong Sun}, {and} \bibinfo{person}{Zhiyuan Liu}.}
  \bibinfo{year}{2018}\natexlab{}.
\newblock \showarticletitle{Entity-duet neural ranking: Understanding the role
  of knowledge graph semantics in neural information retrieval}.
\newblock \bibinfo{journal}{\emph{arXiv preprint arXiv:1805.07591}}
  (\bibinfo{year}{2018}).
\newblock


\bibitem[\protect\citeauthoryear{Luo, Liu, Yang, Bo, Cao, Wu, Li, Yang, and
  Zhu}{Luo et~al\mbox{.}}{2020}]%
        {luo2020alicoco}
\bibfield{author}{\bibinfo{person}{Xusheng Luo}, \bibinfo{person}{Luxin Liu},
  \bibinfo{person}{Yonghua Yang}, \bibinfo{person}{Le Bo},
  \bibinfo{person}{Yuanpeng Cao}, \bibinfo{person}{Jinghang Wu},
  \bibinfo{person}{Qiang Li}, \bibinfo{person}{Keping Yang}, {and}
  \bibinfo{person}{Kenny~Q Zhu}.} \bibinfo{year}{2020}\natexlab{}.
\newblock \showarticletitle{AliCoCo: Alibaba E-commerce Cognitive Concept Net}.
  In \bibinfo{booktitle}{\emph{Proceedings of the 2020 ACM SIGMOD International
  Conference on Management of Data}}. \bibinfo{pages}{313--327}.
\newblock


\bibitem[\protect\citeauthoryear{Mustafa, Khan, and Latif}{Mustafa
  et~al\mbox{.}}{2008}]%
        {mustafa2008ontology}
\bibfield{author}{\bibinfo{person}{Jibran Mustafa},
  \bibinfo{person}{Sharifullah Khan}, {and} \bibinfo{person}{Khalid Latif}.}
  \bibinfo{year}{2008}\natexlab{}.
\newblock \showarticletitle{Ontology based semantic information retrieval}. In
  \bibinfo{booktitle}{\emph{2008 4th International IEEE Conference Intelligent
  Systems}}, Vol.~\bibinfo{volume}{3}. IEEE, \bibinfo{pages}{22--14}.
\newblock


\bibitem[\protect\citeauthoryear{Pang, Knight, and Marcu}{Pang
  et~al\mbox{.}}{2003}]%
        {pang2003syntax}
\bibfield{author}{\bibinfo{person}{Bo Pang}, \bibinfo{person}{Kevin Knight},
  {and} \bibinfo{person}{Daniel Marcu}.} \bibinfo{year}{2003}\natexlab{}.
\newblock \showarticletitle{Syntax-based alignment of multiple translations:
  Extracting paraphrases and generating new sentences}. In
  \bibinfo{booktitle}{\emph{Proceedings of the 2003 Human Language Technology
  Conference of the North American Chapter of the Association for Computational
  Linguistics}}. \bibinfo{pages}{181--188}.
\newblock


\bibitem[\protect\citeauthoryear{Ravichandran and Hovy}{Ravichandran and
  Hovy}{2002}]%
        {ravichandran2002learning}
\bibfield{author}{\bibinfo{person}{Deepak Ravichandran} {and}
  \bibinfo{person}{Eduard Hovy}.} \bibinfo{year}{2002}\natexlab{}.
\newblock \showarticletitle{Learning surface text patterns for a question
  answering system}. In \bibinfo{booktitle}{\emph{Proceedings of the 40th
  Annual meeting of the association for Computational Linguistics}}.
  \bibinfo{pages}{41--47}.
\newblock


\bibitem[\protect\citeauthoryear{Sun, Wang, Li, Feng, Chen, Zhang, Tian, Zhu,
  Tian, and Wu}{Sun et~al\mbox{.}}{2019}]%
        {ernie_baidu}
\bibfield{author}{\bibinfo{person}{Yu Sun}, \bibinfo{person}{Shuohuan Wang},
  \bibinfo{person}{Yu{-}Kun Li}, \bibinfo{person}{Shikun Feng},
  \bibinfo{person}{Xuyi Chen}, \bibinfo{person}{Han Zhang},
  \bibinfo{person}{Xin Tian}, \bibinfo{person}{Danxiang Zhu},
  \bibinfo{person}{Hao Tian}, {and} \bibinfo{person}{Hua Wu}.}
  \bibinfo{year}{2019}\natexlab{}.
\newblock \showarticletitle{{ERNIE:} Enhanced Representation through Knowledge
  Integration}.
\newblock \bibinfo{journal}{\emph{CoRR}}  \bibinfo{volume}{abs/1904.09223}
  (\bibinfo{year}{2019}).
\newblock
\showeprint[arxiv]{1904.09223}
\urldef\tempurl%
\url{http://arxiv.org/abs/1904.09223}
\showURL{%
\tempurl}


\bibitem[\protect\citeauthoryear{Sun, Yang, Zhang, Bozzon, Huang, and Xu}{Sun
  et~al\mbox{.}}{2018}]%
        {sun2018recurrent}
\bibfield{author}{\bibinfo{person}{Zhu Sun}, \bibinfo{person}{Jie Yang},
  \bibinfo{person}{Jie Zhang}, \bibinfo{person}{Alessandro Bozzon},
  \bibinfo{person}{Long-Kai Huang}, {and} \bibinfo{person}{Chi Xu}.}
  \bibinfo{year}{2018}\natexlab{}.
\newblock \showarticletitle{Recurrent knowledge graph embedding for effective
  recommendation}. In \bibinfo{booktitle}{\emph{Proceedings of the 12th ACM
  Conference on Recommender Systems}}. \bibinfo{pages}{297--305}.
\newblock


\bibitem[\protect\citeauthoryear{Sutskever, Vinyals, and Le}{Sutskever
  et~al\mbox{.}}{2014}]%
        {Sutskever2014}
\bibfield{author}{\bibinfo{person}{Ilya Sutskever}, \bibinfo{person}{Oriol
  Vinyals}, {and} \bibinfo{person}{Quoc~V Le}.}
  \bibinfo{year}{2014}\natexlab{}.
\newblock \showarticletitle{Sequence to Sequence Learning with Neural
  Networks}.
\newblock In \bibinfo{booktitle}{\emph{Advances in Neural Information
  Processing Systems 27}}, \bibfield{editor}{\bibinfo{person}{Z.~Ghahramani},
  \bibinfo{person}{M.~Welling}, \bibinfo{person}{C.~Cortes},
  \bibinfo{person}{N.~D. Lawrence}, {and} \bibinfo{person}{K.~Q. Weinberger}}
  (Eds.). \bibinfo{publisher}{Curran Associates, Inc.},
  \bibinfo{pages}{3104--3112}.
\newblock


\bibitem[\protect\citeauthoryear{Szpektor, Tanev, Dagan, and Coppola}{Szpektor
  et~al\mbox{.}}{2004}]%
        {szpektor2004scaling}
\bibfield{author}{\bibinfo{person}{Idan Szpektor}, \bibinfo{person}{Hristo
  Tanev}, \bibinfo{person}{Ido Dagan}, {and} \bibinfo{person}{Bonaventura
  Coppola}.} \bibinfo{year}{2004}\natexlab{}.
\newblock \showarticletitle{Scaling web-based acquisition of entailment
  relations}. In \bibinfo{booktitle}{\emph{Proceedings of the 2004 conference
  on empirical methods in natural language processing}}.
  \bibinfo{pages}{41--48}.
\newblock


\bibitem[\protect\citeauthoryear{Wang, Zhang, Zhao, Li, Xie, and Guo}{Wang
  et~al\mbox{.}}{2019b}]%
        {wang2019multi}
\bibfield{author}{\bibinfo{person}{Hongwei Wang}, \bibinfo{person}{Fuzheng
  Zhang}, \bibinfo{person}{Miao Zhao}, \bibinfo{person}{Wenjie Li},
  \bibinfo{person}{Xing Xie}, {and} \bibinfo{person}{Minyi Guo}.}
  \bibinfo{year}{2019}\natexlab{b}.
\newblock \showarticletitle{Multi-task feature learning for knowledge graph
  enhanced recommendation}. In \bibinfo{booktitle}{\emph{The World Wide Web
  Conference}}. \bibinfo{pages}{2000--2010}.
\newblock


\bibitem[\protect\citeauthoryear{Wang, He, Cao, Liu, and Chua}{Wang
  et~al\mbox{.}}{2019a}]%
        {wang2019kgat}
\bibfield{author}{\bibinfo{person}{Xiang Wang}, \bibinfo{person}{Xiangnan He},
  \bibinfo{person}{Yixin Cao}, \bibinfo{person}{Meng Liu}, {and}
  \bibinfo{person}{Tat-Seng Chua}.} \bibinfo{year}{2019}\natexlab{a}.
\newblock \showarticletitle{Kgat: Knowledge graph attention network for
  recommendation}. In \bibinfo{booktitle}{\emph{Proceedings of the 25th ACM
  SIGKDD International Conference on Knowledge Discovery \& Data Mining}}.
  \bibinfo{pages}{950--958}.
\newblock


\bibitem[\protect\citeauthoryear{Wise, Ioannidis, Calvo, Song, Price, Kulkarni,
  Brand, Bhatia, and Karypis}{Wise et~al\mbox{.}}{2020}]%
        {wise2020covid}
\bibfield{author}{\bibinfo{person}{Colby Wise}, \bibinfo{person}{Vassilis~N
  Ioannidis}, \bibinfo{person}{Miguel~Romero Calvo}, \bibinfo{person}{Xiang
  Song}, \bibinfo{person}{George Price}, \bibinfo{person}{Ninad Kulkarni},
  \bibinfo{person}{Ryan Brand}, \bibinfo{person}{Parminder Bhatia}, {and}
  \bibinfo{person}{George Karypis}.} \bibinfo{year}{2020}\natexlab{}.
\newblock \showarticletitle{COVID-19 knowledge graph: accelerating information
  retrieval and discovery for scientific literature}.
\newblock \bibinfo{journal}{\emph{arXiv preprint arXiv:2007.12731}}
  (\bibinfo{year}{2020}).
\newblock


\bibitem[\protect\citeauthoryear{Wu, Li, Wang, and Zhu}{Wu
  et~al\mbox{.}}{2012}]%
        {wu2012probase}
\bibfield{author}{\bibinfo{person}{Wentao Wu}, \bibinfo{person}{Hongsong Li},
  \bibinfo{person}{Haixun Wang}, {and} \bibinfo{person}{Kenny~Q Zhu}.}
  \bibinfo{year}{2012}\natexlab{}.
\newblock \showarticletitle{Probase: A probabilistic taxonomy for text
  understanding}. In \bibinfo{booktitle}{\emph{Proceedings of the 2012 ACM
  SIGMOD International Conference on Management of Data}}.
  \bibinfo{pages}{481--492}.
\newblock


\bibitem[\protect\citeauthoryear{Zhao, Wang, Liu, and Li}{Zhao
  et~al\mbox{.}}{2008}]%
        {zhao2008pivot}
\bibfield{author}{\bibinfo{person}{Shiqi Zhao}, \bibinfo{person}{Haifeng Wang},
  \bibinfo{person}{Ting Liu}, {and} \bibinfo{person}{Sheng Li}.}
  \bibinfo{year}{2008}\natexlab{}.
\newblock \showarticletitle{Pivot approach for extracting paraphrase patterns
  from bilingual corpora}. In \bibinfo{booktitle}{\emph{Proceedings of ACL-08:
  HLT}}. \bibinfo{pages}{780--788}.
\newblock


\end{thebibliography}
\end{CJK}
\end{document}